\documentclass[aps,prb,twocolumn,groupedaddress]{revtex4-1}
\usepackage{graphicx,epsfig,epstopdf}
\usepackage{amsmath}
\usepackage{amssymb}
\usepackage{float}
\usepackage{xspace}
\usepackage{color}
\usepackage{soul}

\newcommand{\SHS}{Shastry-Sutherland }

\newcommand{\SCBO}{S\lowercase{r}C\lowercase{u}$_2$(BO$_3$)$_2$ }
\newcommand{\SCBOO}{S\lowercase{r}C\lowercase{u}$_2$(BO$_3$)$_2$}
\newcommand{\INS}{inelastic neutron scattering }
\newcommand{\IINS}{Inelastic neutron scattering }

\begin{document}

\title{Observation of a 4-spin Plaquette Singlet State in the Shastry--Sutherland compound SrCu$_2$(BO$_3$)$_2$}

\author{
M.E.~Zayed$^{1,2,3}$,
Ch.~R\"uegg$^{3,4}$,
J.~Larrea~J.$^{2,5}$,
A.M.~L\"auchli$^{6}$,
C.~Panagopoulos$^{7,8}$,
S.S.~Saxena$^{7}$,
M.~Ellerby$^{9}$,
D.F.~McMorrow$^{9}$,
Th.~Str\"assle$^{3}$,
S.~Klotz,$^{10}$,
G.~Hamel$^{10}$,
R.A.~Sadykov$^{11,12}$,
V.~Pomjakushin,$^{3}$,
M.~Boehm$^{13}$,
M.~Jim\'enez--Ruiz$^{13}$,
A.~Schneidewind $^{14}$,
E.~Pomjakushina$^{15}$,
M.~Stingaciu$^{15}$,
K.~Conder$^{15}$,
and H.M.~R$\o$nnow$^{2}$}
   
\affiliation{
\\
$^1$ Department of Mathematics, Statistics and Physics, College of Arts and Science, Qatar University, P.O. Box 2713 Doha, Qatar\\
$^2$ Laboratory for Quantum Magnetism, Ecole Polytechnique Federal de Lausanne (EPFL), 1015 Lausanne, Switzerland \\
$^3$ Laboratory for Neutron Scattering and Imaging, Paul Scherrer Institut, 5232 Villigen PSI, Switzerland \\
$^4$ Department of Quantum Matter Physics, University of Geneva, 1211 Geneva 4, Switzerland \\
$^5$ Centro Brasileiro de Pesquisas Fisicas, Rua Doutor Xavier Sigaud 150,CEP 2290-180, Rio de Janeiro, Brazil \\
$^6$ Institut f\"ur Theoretische Physik, Universit\"at Innsbruck, 6020 Innsbruck, Austria \\
$^7$ Cavendish Laboratory, University of Cambridge, Cambridge CB3 0HE, UK \\
$^8$ Division of Physics and Applied Physics, School of Physical and Mathematical Sciences, Nanyang Technological University, Singapore 637371\\
$^9$ London Centre for Nanotechnology and Department of Physics and Astronomy, University College London, London WC1E 6BT, UK \\
$^{10}$ IMPMC; CNRS--UMR 7590, Universite Pierre et Marie Curie, 75252 Paris, France \\
$^{11}$ Institute for Nuclear Research, Russian Academy of Sciences,  prospekt 60-letiya Oktyabrya 7a, Moscow 117312.\\
$^{12}$ Vereshchagin Institute for High Pressure Physics, Russian Academy of Sciences, 142190 Troitsk, Russia\\
$^{13}$ Institut Laue-Langevin, 71 avenue des Martyrs - CS 20156- 38042 Grenoble Cedex 9, France \\
$^{14}$ J\"ulich Centre for Neutron Science (JCNS),
 Forschungszentrum J\"ulich GmbH, Outstation at Heinz Maier-Leibnitz Zentrum (MLZ), Lichtenbergstra{\ss}e 1, D-85747 Garching, Germany\\
$^{15}$ Laboratory for Scientific Developments and Novel Materials, Paul Scherrer Institut, 5232 Villigen PSI, Switzerland}

\date{\today}

\maketitle

{\bf The study of interacting spin systems is of fundamental importance for modern condensed matter physics. On frustrated lattices, magnetic exchange interactions cannot be simultaneously satisfied, and often give rise to competing exotic ground states\cite{Lhuillier01}. The frustrated 2D Shastry-Sutherland lattice\cite{Shastry81} realized by \SCBO\cite{Miyahara03R} is an important test to our understanding of quantum magnetism. It was constructed to have an exactly solvable 2-spin dimer singlet ground state within a certain range of exchange parameters and frustration. While the exact dimer state and the antiferromagnetic order at both ends of the phase diagram are well known, the ground state and spin correlations in the intermediate frustration range have been widely debated \cite{Miyahara03R,Shastry81,Albrecht96,Weihong99,Muller00,Knetter00a,Koga00a,Takushima01,Weihong02,Lauchli02,AlHajj05}. We report here the first experimental identification of the conjectured plaquette singlet intermediate phase in \SCBOO. It is observed by \INS after pressure tuning at 21.5~kbar. This gapped plaquette singlet state with strong 4-spin correlations leads to a transition to an ordered N\'eel state above 40~kbar, which can realize a deconfined quantum critical point.}\\

In the field of quantum magnetism, geometrically
frustrated lattices generally imply major difficulties in analytical
and numerical studies. For very few particular topologies however, it
has been shown that the ground state, at least, can be calculated
\textit{exactly} as for the Majumdar-Gosh model\cite{Majumdar69} that
solves the $J_1$-$J_2$ zig-zag chain when $J_1=2J_2$. In 2D, the
Shastry-Sutherland model \cite{Shastry81} consisting of an
orthogonal dimer network of spin S=1/2 was developed in order to be
exactly solvable. For an inter-dimer $J^{\prime}$ to intra-dimer $J$ exchange ratio
$\alpha\equiv J^{\prime}/J\leq0.5$ the ground state is a product of
singlets on the strong bond $J$. Numerical calculations have further
shown that this remains valid up to $\alpha\leq\sim0.7$ and for
small values of 3D couplings $J^{\prime\prime}$ between dimer layers. 
At the other end,
for $\sim0.9\leq\alpha\leq\infty$ 
the system approaches the well known 2D square lattice, which is antiferromagnetically (AFM) ordered, albeit with significant quantum fluctuations that are believed to include resonating singlet correlations resulting in fractional excitations\cite{DallaPiazza15}.
The phase diagram of the \SHS model,
both with and without applied magnetic field, has been intensively studied by
numerous theoretical and numerical approaches\cite{Miyahara03R}. In the presence of magnetic field,
magnetization plateaus at fractional values of the saturation magnetization corresponding to Mott insulator phases of dimer states, as well as possible superfluid and supersolid phases have been extensively studied\cite{Muller00,Momoi00b,Dorier08}.
At zero field, the main unsolved issue is the existence and nature of an
intermediate phase for $\sim0.7\leq\alpha\leq\sim0.9$. A variety of quantum phases and transitions between them
have been predicted depending on the theoretical technique used: a direct transition
from dimer singlet phase to AFM order\cite{Shastry81,Muller00,Weihong99},
or an intermediate phase with helical order\cite{Albrecht96}, columnar
dimers\cite{Weihong02}, valence bond crystal\cite{Lauchli02} or
resonating valence bond (RVB) plaquettes\cite{Koga00a,Takushima01}. Recent
results indicate that a plaquette singlet phase is favored \cite{Miyahara03R,Corboz13}. From such
a phase, which would have an additional Ising-type order parameter, a subsequent transition to AFM order
could provide a realisation of the so far elusive deconfined quantum critical point \cite{Senthil04}.

The compound strontium copper borate \SCBO is the only known realization of the \SHS model with S=1/2
spins\cite{Miyahara03R} and has thus triggered considerable attention in
the field of quantum magnetism. The spectrum of
SrCu$_2$(BO$_3$)$_2$ exhibits an almost dispersionless $\Delta$=3~meV gap, and a bound-state of two
triplets (BT) forms at $ E_{BT}\simeq$5~meV. The unusual size and dispersionless nature of the gap is an
effect of the frustration which prevents triplets from hopping up to sixth order \cite{Miyahara03R}. The estimated exchange parameters
in the material $J\sim85$~K and $\alpha =0.635$ \cite {Miyahara03R} or $J\sim71$~K and
$\alpha =0.603$ \cite{Knetter00a} place the compound close to an interesting regime $\alpha\sim0.7$ where correlations
may change dramatically at a critical point.

A precious mean to tune a quantum magnet across a quantum phase transition is the
application of hydrostatic pressure as it directly modifies the
atomic distances and bridging angles, such as Cu-O-Cu and thus the magnetic exchange
integrals. Quantum phase transitions were successfully discovered
in dimer magnets upon application of pressure \cite{Merchant14}. However
high pressure measurements remain technically challenging. In the
case of SrCu$_2$(BO$_3$)$_2$ magnetic susceptibility\cite{Kageyama03} and ESR\cite{Sakurai09} to moderate
pressures (p$\leq$12~kbar) indicate a softening of the gap, while the
combined effect of pressure and field was measured by susceptibility
and NMR\cite{Waki07}. In the latter case magnetic order occurring at
24~kbar and 7~T on a fraction of the dimers was proposed. 
In an X-ray diffraction investigation the temperature dependence of the lattice parameters was analysed as
an indirect proxy for the singlet triplet gap leading to the suggestion that it closes at 20~kbar \cite{Haravifard12}.
At even higher pressures neutron and X-ray diffraction experiments observed a transition  above 45~kbar from the ambient I$\bar{4}$2M
tetragonal space group to monoclinic \cite{Loa05,Zayed14,Haravifard14}.

\begin{figure}[t]
\includegraphics[width=\columnwidth]{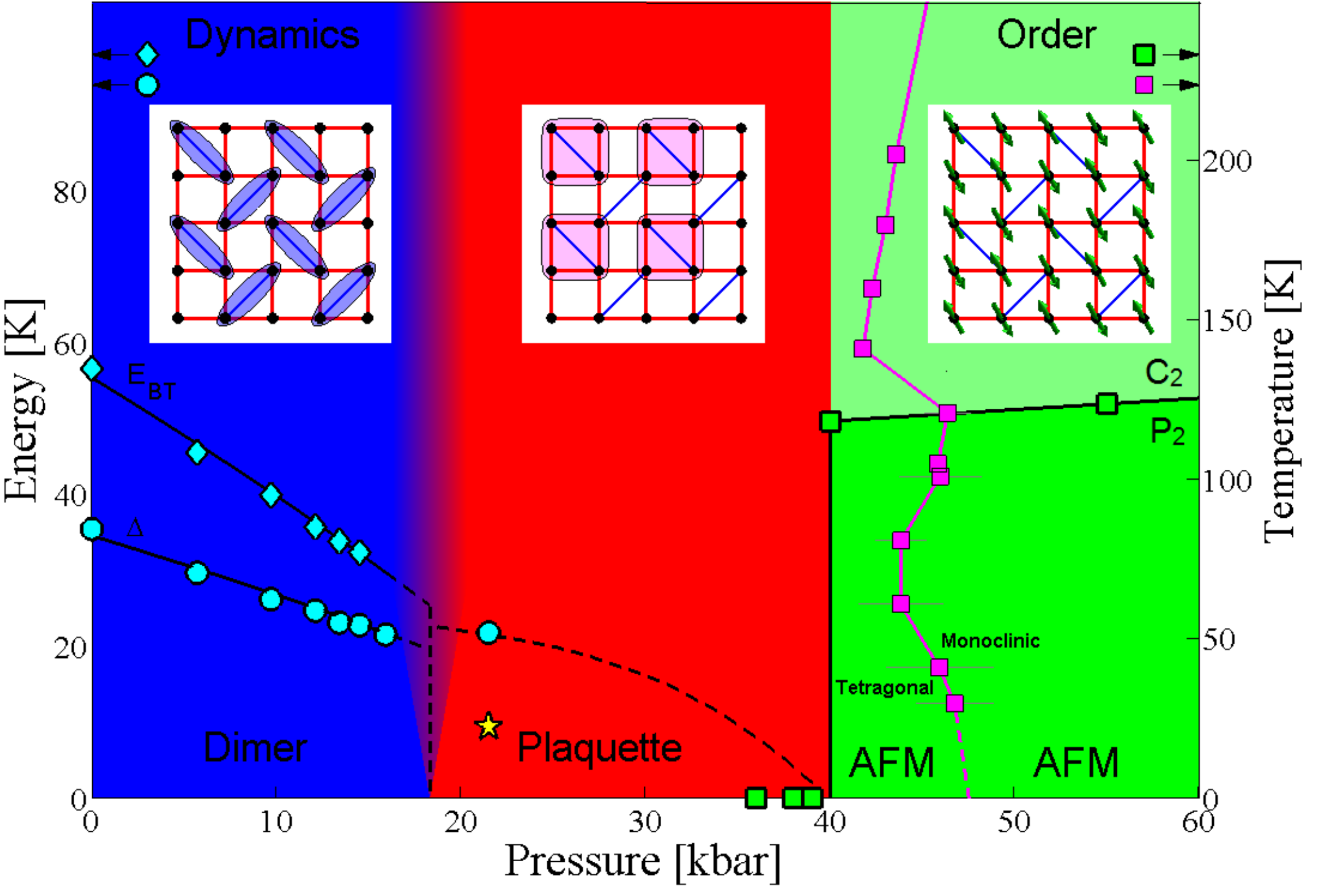}
\caption{Phase diagram of \SCBO as a function of pressure and temperature, including excitation energies. The blue region is the dimer phase, the red region the newly identified plaquette phase, and the green region the antiferromagnetic phases where  Q=(1,0,0) magnetic Bragg peaks, indicated by green squares, are observed only above 40~kbar. Circles are the triplet gap energy $\Delta$ at Q=(2,0,L), diamonds are the corresponding 2-triplet bound state energy E$_{BT}$ and the star is a new low energy excitation observed at Q=(1,0,1). The magenta line shows the tetragonal to monoclinic structural transition \cite{Zayed14}. The corresponding monoclinic space groups are indicated \cite{Zayed10T, Haravifard14}. The dashed line in the plaquette phase is the extrapolated energy gap using Ref.~\onlinecite{Koga00a}. The insets depict the corresponding ground states. All the experimental points are from this study.}
\end{figure}

Here we present neutron spectroscopy results, which directly determine the pressure dependence of the gap and through the dynamic structure factor allows us to address the nature of the correlations. Figure~1 summarize the phase diagram of \SCBOO, we determined in this study. The exact dimer phase survives up to 16~kbar. The gap decreases from 3~meV to 2~meV, but does not vanish. 
At 21.5~kbar we discover experimentally a new, intermediate phase. 
We identify it by its \INS spectrum as the formation of 4-spin plaquette singlets.
Above 40~kbar and below 117~K we find by neutron diffraction that AFM order appears while the compound likely still has tetragonal symmetry with orthogonal dimers. Above $\sim$45~kbar, a structural distortion takes place and the symmetry becomes monoclinic, implying non-orthogonal dimers \cite{Zayed14,Haravifard14}. \SCBO is magnetically ordered after the distortion, but can no longer be described appropriately by the original \SHS model. 
The transition from 2-spin dimer to 4-spin plaquette singlets appears to be of first order, whereas the transition from the plaquette to the AFM phase could be of second order and concomitant with the continious closure of the plaquette gap or of first order~\cite{Koga00a,Koga00c,Block13,Corboz13}.

To allow a quantitative comparison to theoretical predictions we establish the pressure dependence of the exchange parameters $J_{\chi}(p)$, $J^{\prime}_{\chi}(p)$, and $\alpha(p)$ by measuring magnetic susceptibility $\chi(p,T)$ and fitting it using 20 sites exact-diagonalization.
The peak in susceptibility shifts to lower temperature as pressure increases up to 10~kbar (Figure~2a). This suggest a reduction of the spin gap. We parametrise the pressure dependence of $J$ and $J^{\prime}$ by linear fits (Figure~2b). $J$ has the larger slope so that $\alpha$ increases with pressure.
Having established $\alpha(p)$ we see that the critical pressure lying between 16~kbar and 21.5~kbar corresponds to $0.66<\alpha_{c}<0.68$, in good agreement with theoretical predictions\cite{Miyahara03R,Lauchli02,Corboz13}.

\begin{figure}[t]
\includegraphics[width=\columnwidth]{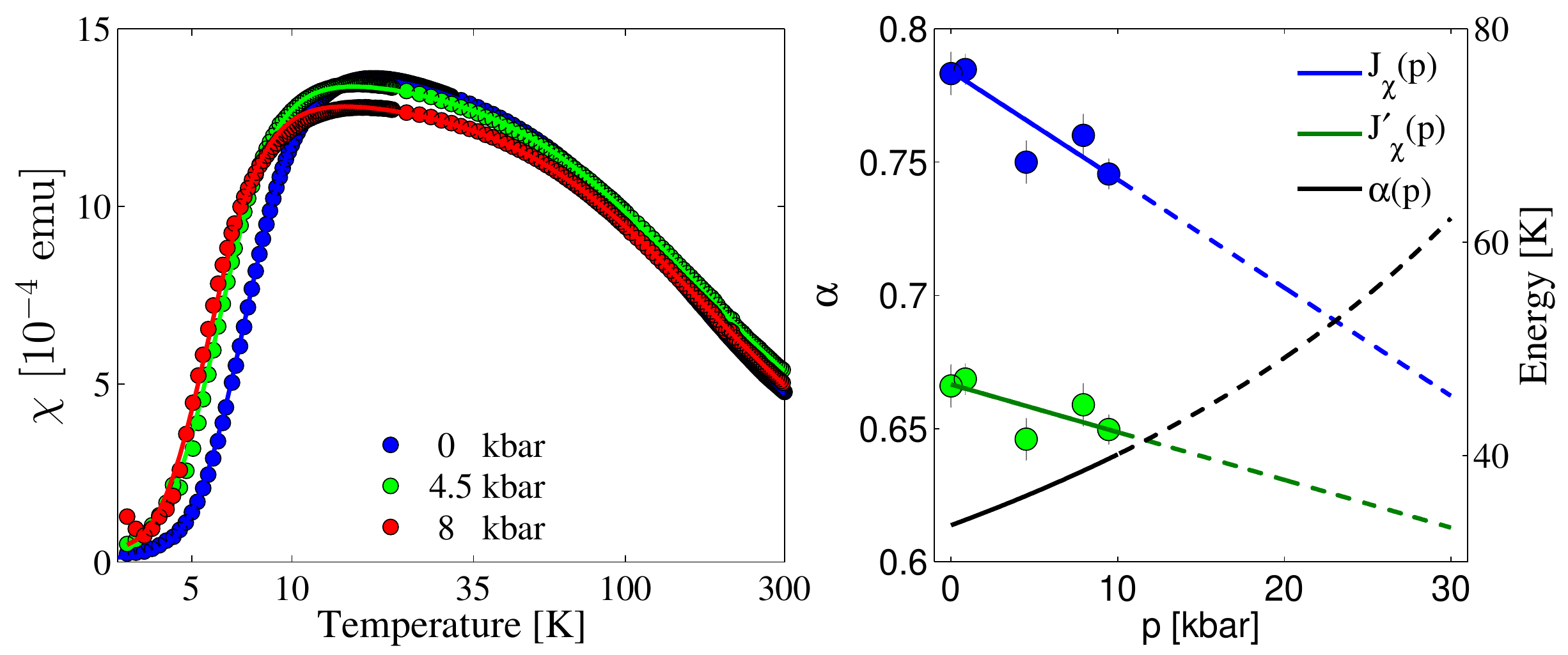}
\caption{Pressure dependence of the magnetic susceptibility and of the exchange parameters in \SCBOO. Left, magnetic susceptibility at three pressures below 10~kbar with fits to calculations by exact diagonalization (solid lines), $H$=0.5~T. Right, extracted exchange parameters $J_{\chi}(p)$ and $J^{\prime}_{\chi}(p)$ with linear fits and their ratio $\alpha(p)$.}
\end{figure}

A selection from the neutron spectra leading to the phase diagram are summarised in Figure~3.
Up to 16~kbar an
essentially Q-independent linear decrease of the gap energy is
observed (Figures 1 and 3a). The measurement of the dispersion and of
the structure factor in that pressure range shows that the spin system is
still in its original "exact dimer" ground-state with a reduced
energy scale of the exchange parameters.
The dispersion increases slightly with pressure (not shown),
which can be understood by the increase of
$\alpha$~\cite{Weihong99}. 
Interestingly, the bound triplet energy $E_{BT}$ softens twice
as fast, implying that the triplet binding
energy, $\delta=2\Delta-E_{BT}$=1.19(2)~meV, remains pressure
independent.
This results in the unusual situation that extrapolating the softenings, the bound triplet
would reach zero energy before the single triplet, and hence that, before that point, exciting a bound state of two triplets would cost less energy than exciting one triplet.

\begin{figure}[t]
\includegraphics[width=\columnwidth]{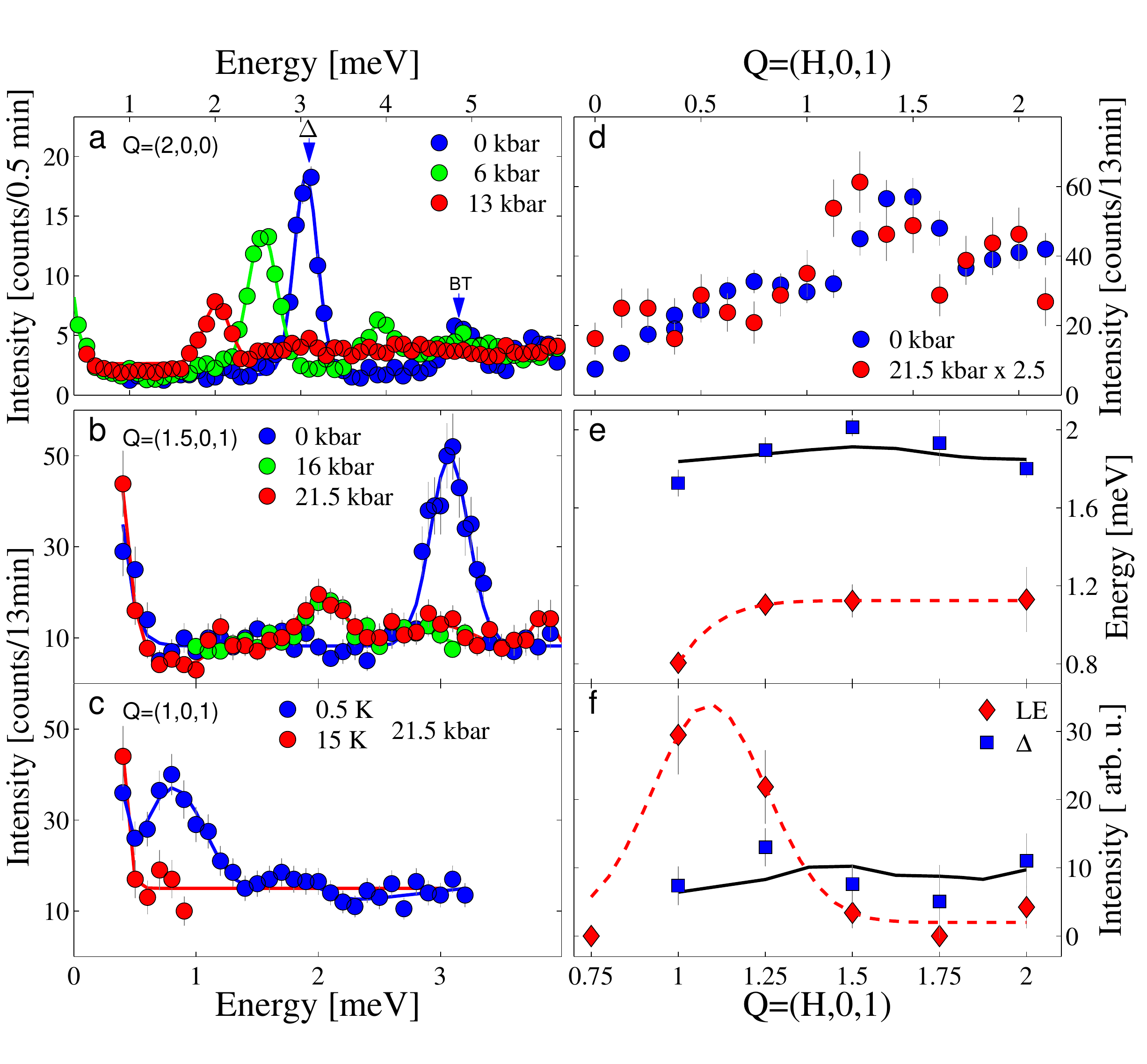}
\caption{\IINS measurements of \SCBO under hydrostatic pressure. (a) Energy spectra with triplet gap $\Delta$ and 2-triplet bound state E$_{BT}$ energies softening in the dimer phase (setup 1). (b) Discontinuity in the gap softening at 16 and 21.5~kbar (setup 2). (c) New low energy excitation LE at Q=(1,0,1) (setups 2,3). (d) Momentum dependence at the gap energy $\Delta$ (setup 2). (e-f) Dispersion and intensity for the triplet $\Delta$ and LE, the solid lines are scaled ambient pressure values adapted from\cite{Kakurai05} and the dashed lines are guides to the eye.}
\end{figure}

\SCBO enters a new quantum phase between 16 and 21.5~kbar,
where a discontinuity in the gap softening occurs.
The spectra at these two pressures for Q=(1.5,0,1) (Figure 3b)  and for (2,0,0) are extremely similar and the gap energy
$\Delta\simeq$2~meV remains unchanged.
The transition to a new quantum phase is further asserted by a new
type of excitation suddenly appearing at the higher pressure (Figure 3c).
It is clearly visible around 1~meV for Q=(1,0,1), (-1,0,1) and
(1,0,1.5) at 0.5~K and is not observed at 15~K, which proves the
magnetic origin. Figures 3e and 3f show the dispersion and intensity
of the two excitations along (1$\leq$H$\leq$2,0,1). The excitation at 2~meV
clearly displays a similar behaviour as that expected from
the singlet-triplet gap excitation in the exact dimer phase (given by
the full lines) and we thus keep labelling it $\Delta$. The new low
energy excitation (LE) on the other hand is more dispersive,
$\sim$0.4~meV in the measured momentum range, and has a different structure
factor strongly peaking at Q=(1,0,1). Another strong evidence for
the similarity of $\Delta$ and the ambient pressure gap is given by
the comparison of scans at constant energy along (H,0,1) at 0 and 21.5~kbar in Figure
3d. Except for the overall decrease in intensity, the two scans are
close to identical.

\begin{figure}[t]
\includegraphics[width=\columnwidth]{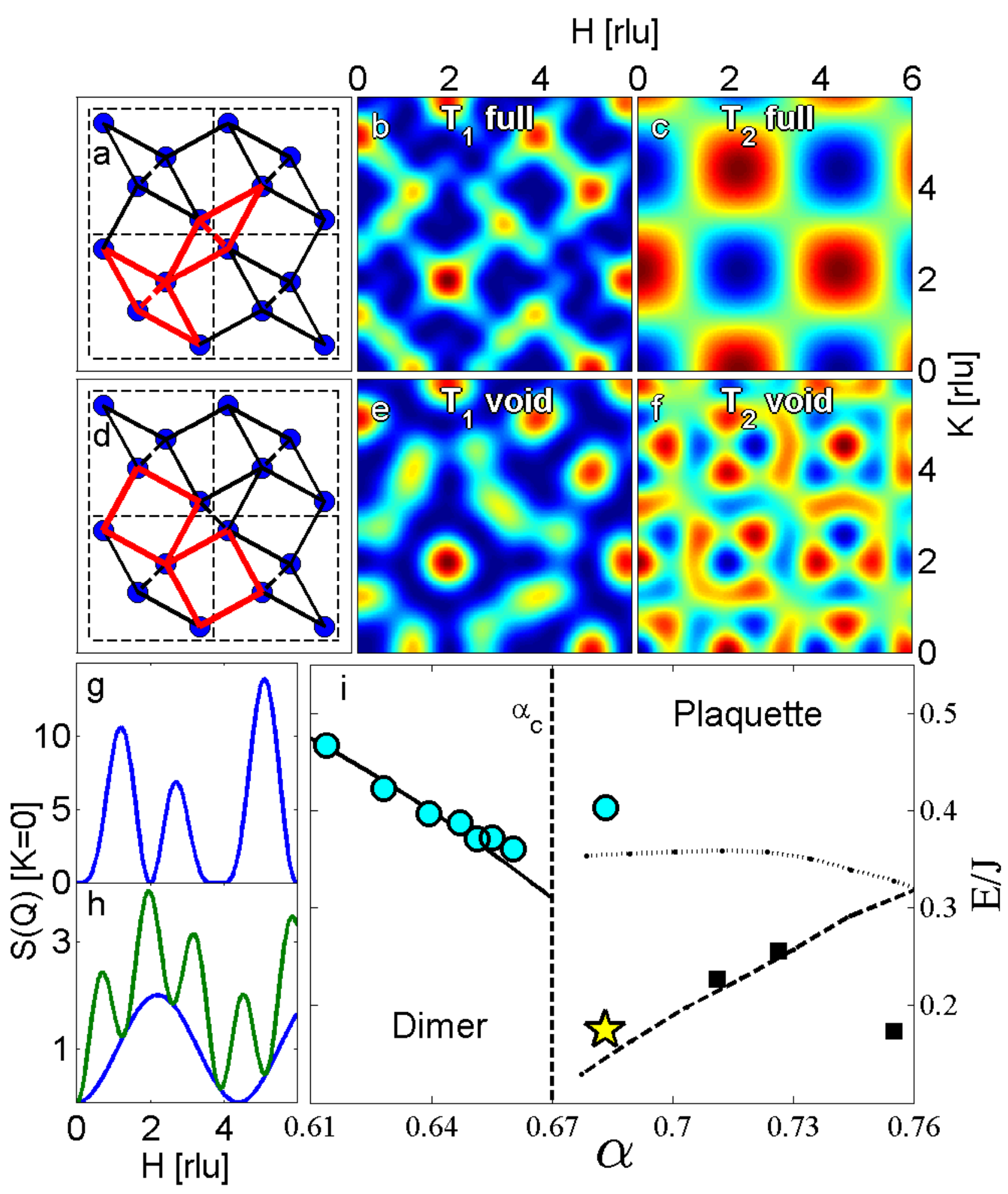}
\caption{Plaquette phase in \SCBOO. (a,d) Plaquettes containing a diagonal bound (full plaquettes) and void plaquettes. The structure factors $S^{xx}$+$S^{yy}$+$S^{zz}$ for T$_1$ (b,e) and T$_2$ (c,f) are calculated as the sum over the two possible plaquette orientations in the \SCBO geometry and (c) is identical to the structure factor of two orthogonal isolated dimers. (g) Structure factor along Q$_{hk}$=(Q$_h$,Q$_k$)=(H,0) for T$_1$. Void and full plaquette are identical. (h) Structure factor along Q$_{hk}$=(H,0) for T$_2$, void plaquette in green and full plaquette in blue. The blue line is also the isolated dimer structure factor.
(i) Comparison of excitation energies between experiment (same points as in Fig. 1) and theory calculations: dimer gap energy adapted from Ref.~\onlinecite{Weihong02} (full line), low and high energy triplet excitations in the plaquette phase from Ref.~\onlinecite{Takushima01} (dotted lines), and for columnar plaquette block energies Ref.~\onlinecite{AlHajj05} (black squares).}
\end{figure}

To interpret the appearance of a new excitation and the observed momentum dependence of the dynamical structure factors, it is illustrative to consider the simplified case of an isolated 4-spin plaquette, described in the Methods section, which has a singlet ground state and shows two low lying excitations T$_1$ and T$_2$. The structure factors of these excitations, summed over the two possible 'full' plaquette
orientations (Figure 4a), are shown in Figures 4b-c together with those of a 'void' plaquette (Figures 4d-f) containing
no diagonal bond. T$_1$ has a structure factor peaking near Q$_{hk}$=(Q$_h$,Q$_k$)=(1,0) in the 2D geometry
of \SCBO for both full and void plaquettes (Figure 4g). T$_2$, however,
has a structure factor identical to that of an isolated dimer on the diagonal bond
only for the full plaquette (Figures 4c, 4f, and 4h).
While an extended many-body calculation would be needed for a fully quantitative comparison, the isolated plaquette considered here
displays the main characteristics of the new intermediate pressure phase: (1) a non-magnetic gapped ground state,
(2) a low energy triplet (LE) with structure factor peaking at Q$_{hk}$=(1,0), and (3) another low energy
excitation ($\Delta$) with structure factor identical to the singlet-triplet transition in the exact dimer phase.
We thus identify the discovered phase as composed of 4-spin plaquette singlets, with excitation LE corresponding to T$_1$ and excitation $\Delta$ corresponding to T$_2$.
Comparing the experimental intensities to this simple calculation favors the singlets sitting on 'full' plaquettes containing diagonal bonds, but calculations of the structure factor for the extended model are required for verification of this point.\\

To analyze further the interacting plaquette system, we plot in Figure 4i the measured energies $E/J$
vs. $\alpha$ which enables a direct comparison between our results and the calculations for the
low- and high-energy RVB-like plaquette excitations by Ref.~\onlinecite{Takushima01}, and
columnar plaquette block energy~\cite{AlHajj05}. Experimental and calculated \cite{Weihong99} gap
energies in the dimer phase are in excelent agreement. Beyond the transition, there is qualitative
agreement for the energy scales, in particular the observed energies of LE and of $\Delta$ for 21.5~kbar are close to the expected low- and high-energy plaquette excitations of Ref.~\onlinecite{Takushima01} for $\alpha=0.68$. 

Our results can also explain the occurrence of magnetic ordering proposed by NMR measurements at 24~kbar and 7~T \cite{Waki07}:
the new spin S=1 excitation LE being low in energy (0.5~meV), a 7~T field is
sufficient to close the related gap and to obtain a magnetic ground state. This field-induced
quantum critical point and resulting phase will be related to the field-induced BEC physics observed in
dimer singlet systems \cite{Ruegg03}, but could reveal new phenomena due to the strong frustration in
the \SHS model. Especially, the evolution of the magnetisation plateaus in \SCBO
with pressure remains to be studied. Based on our results presented here we can predict that
in particular the pressure range between 15 and 25~kbar will be of high interest.\\

In conclusion we have performed high pressure experiments on
SrCu$_2$(BO$_3$)$_2$ and tuned the compound to experimentally
identify a plaquette singlet phase at intermediate exchange ratio in the
Shastry-Suterland lattice. We observed a first order transition taking place
between two magnetically disordered states: the exact 2-spin dimer singlet and the 4-spin plaquette
singlet phase. The dominant correlations in the plaquette phase involve a four-spin unit
and are characterised by a low-lying triplet excitation
that is not present in the dimer phase and that gives access to new types of field- and
pressure-induced quantum critical points.
The plaquette phase itself is suppressed at higher pressures were
classical N\'eel order is found.
Particularly exciting is the fact that the existence of two possible plaquette singlet coverings offer an Ising-type order parameter. This may turn the transition from plaquette to N\'eel phase into a deconfined quantum critical point at 40~kbar.

\bibliographystyle{naturemag}

{\small
\section{Methods} \noindent

{\bf Experiments.} We measured the excitation spectrum of SrCu$_2$(BO$_3$)$_2$ by
inelastic neutron scattering (INS) on respectively 3~g single
crystals up to 15~kbar and 0.2~g at 16 and 21.5~kbar with piston cylinder
clamped pressure cells with different experimental setups:
Setup~1: IN14 ILL, k$_f$=1.3~$\AA^{-1}$, aluminium-steel pressure cell, p$\leq$17~kbar (HPCAL17),  He-pumped
cryostat, T=1.5~K. Setup~2: IN14, ILL, k$_f$=1.5~$\AA^{-1}$, McWhan pressure cell, p$\leq$22~kbar, He-direct flow cryostat, T=2~K. Setup~3: Same as setup 2, with $^3$He cryostat, T=0.5~K. High pressure INS data was also collected with the following setups: Setup~4: TASP, SINQ-PSI, k$_f$=1.3~$\AA^{-1}$, aluminum pressure cell, p$\leq$12~kbar. Setup~5:  PANDA, FRM-2, k$_f$=1.5~$\AA^{-1}$,  HPCAL17. Setup~4-5 have He-pumped
cryostats, T=1.5~K.
Given the unusual temperature dependence in
SrCu$_2$(BO$_3$)$_2$, where the INS intensity of the 35~K gap is
reduced by half already around 7~K\cite{Zayed14b}, we
performed the measurements at 21.5~kbar both at 2~K and 0.5~K but no significant change in the
excitation intensity was observed. AFM order was observed by neutron diffraction on IN8, ILL,
k$_f$=2.66$\AA^{-1}$, p$\leq$65~kbar with a  Paris-Edinburgh press. Pressure was determined by the shift in lattice constant of Pb or NaCl reference to $\sim$0.7~kbar accuracy.
The pressure dependence of magnetic susceptibility was measured on a MPMS SQUID magnetometer (Quantum Design)
using non-magnetic CuBe clamp pressure cells (CamCell). Pressure was calibrated by the superconducting transition of Pb.\\

\noindent {\bf Data analysis.} The pressure dependent gap $\Delta(J_{\chi}(p), J^{\prime}_{\chi}(p))$ obtained
through the Q=0 expansion of Ref.~\onlinecite{Weihong02} with exchange parameters from fits to susceptibility data
is in good agreement with the direct INS gap measurement
$\Delta_Q(p)$. To take into account the small Q-dependence of
$\Delta_Q$, due to Dzyaloshinskii-Moriya interactions\cite{Kakurai05},
we additionally used $\Delta_Q(p) = \Delta(J_{\chi}(p),J^{\prime}_{\chi}(p)) + D_Q(p)$, where the
correction factor $D_Q$ is of the order of 0.2~meV.\\

\noindent The 4-spin plaquette is described by the Hamiltonian:
\begin{equation}
\mathcal{H}=J^{\prime}(\vec{S}_1\vec{S}_2+\vec{S}_2\vec{S}_3+\vec{S}_3\vec{S}_4+\vec{S}_1\vec{S}_4)+J(\vec{S}_1\vec{S}_3),
\end{equation}
\noindent where the last term represents a diagonal bond between sites 1 and 3 (a 'full' plaquette), and should be removed for a 'void' plaquette without such a diagonal bond. The eigenstates of $\mathcal{H}$ can be separated over two sectors depending on the
value of the quantum number S$_{1,3}$ for the spins $\vec{S_1}+\vec{S_3}$
on the diagonal bond and S$_{2,4}$ for the spins $\vec{S_2}+\vec{S_4}$ on
the outer sites\cite{Koga00c,Zayed10T}. A study of the excitation spectrum
of such a plaquette shows that for $\alpha\geq 0.5$ the ground state is an S=0
singlet of four spins. Two low-lying excitations T$_1$ and T$_2$ are present.
For $\alpha\geq 1$, T$_1$ has the lower energy, while for $0.5 \geq \alpha\geq 1$ T$_2$ does.
T$_1$ corresponds to a triplet excitation with both S$_{1,3}$ and S$_{2,4}$ equal to 1.
In the full plaquette, T$_2$ is four-fold degenerate and corresponds to a singlet on the diagonal
S$_{1,3}$=0 plus two free spins, S$_{2,4}$= 0 or 1. The corresponding structure factor is identical to that of the singlet-triplet excitation on the isolated diagonal bond. For the void plaquette, T$_2$ is sevenfold degenerate and the structure factor does not match the isolated dimer.
\\
\\
\section{Acknowledgments}
We thank A. Magee for her contributions to the susceptibility
measurements, M. Merlini and M. Hanfland for support during
high-pressure X-ray diffraction experiments at the ESRF, and
M. Ay and P. Link for assistance during neutron scattering experiments.
We acknowledge F. Mila and B. Normand for many useful discussions.
This work is based on experiments performed at the Swiss spallation neutron
source SINQ, Paul Scherrer Institute, Villigen, Switzerland,
at the FRM-2, Munich, Germany, and at the ILL, Grenoble, France.
We thank the Swiss National Science Foundation SNF and the Royal Society (UK) for financial support.
JLJ also acknowledges CNPq/MCTI.
\\
\\
\section{Author contribution}
M.E.Z, C.R and H.M.R designed the research, performed the experiments and analyzed the data. A.L. computed the magnetic susceptibility by exact diagonalization.  C.P, S.S.S and M.E helped with susceptibility experiments. S.S.S, T.S, S.K, G.H and R.A.S provided neutron high pressure techniques. M.B, M.J.R, A.S, V.P. and T.S. provided support for neutron experiments.
E.P, M.S., and K.C. synthesized the \SCBO samples. J.L.J. and D.M. contributed to interpretation of the data. M.E.Z, C.R and H.M.R wrote the manuscript with contributions from all co-authors.
}

\end{document}